\documentclass[12pt]{article}
\usepackage{graphicx}

\usepackage{subfig}

\newcommand\pubnumber{
CERN-TH-2017-253\\
HERWIG-2017-03\\
IPPP/17/93\\
KA-TP-41-2017\\
LU-TP-17-41\\
MCnet-17-21\\
UWTHPH-2017-42\\
}
\newcommand\pubdate{\today}

\def\institute{
  $^{1}$Theoretical Particle Physics, Department of Astronomy and Theoretical Physics,
  Lund University, Lund, Sweden\\
  $^{2}$Department of Physics, University of Toronto, Toronto, Canada\\
  $^{3}$Institute for Theoretical Physics, Karlsruhe Institute of Technology, Karlsruhe, Germany\\
  $^{4}$Particle Physics, Faculty of Physics, University of Vienna, Vienna, Austria\\
  $^{5}$HEP Theory Group, Department of Physics, Florida State University, Florida, US\\
  $^{6}$Theoretical Physics Department, CERN, Geneva, Switzerland\\
  $^{7}$IPPP, Department of Physics, Durham University, Durham, UK
}

\def\thefootnote{\fnsymbol{footnote}}
\def\support{\footnote[1]{Speaker}}

\def\Title#1{\begin{center} {\Large #1 } \end{center}}
\def\Author#1{\begin{center}{ \sc #1} \end{center}}
\def\Address#1{\begin{center}{ \it #1} \end{center}}

\newcommand\pubblock{\rightline{\begin{tabular}{l} \pubnumber\\
         \pubdate  \end{tabular}}}
\newenvironment{Abstract}{\begin{quotation}  }{\end{quotation}}
\newenvironment{Presented}{\begin{quotation} \begin{center} 
             PRESENTED AT\end{center}\bigskip 
      \begin{center}\begin{large}}{\end{large}\end{center} \end{quotation}}
\def\Acknowledgements{\bigskip  \bigskip \begin{center} \begin{large}
             \bf ACKNOWLEDGEMENTS \end{large}\end{center}}

\def\beq{\begin{equation}}
\def\eeq#1{\label{#1}\end{equation}}
\def\eeqn{\end{equation}}

\def\beqa{\begin{eqnarray}}
\def\eeqa#1{\label{#1}\end{eqnarray}}
\def\eeqan{\end{eqnarray}}

\let\bar=\overbar

\def\Dslash{\not{\hbox{\kern-4pt $D$}}}
\def\dslash{\not{\hbox{\kern-2pt $\del$}}}

\def\msb{{\bar{\ssstyle M \kern -1pt S}}}

\newcommand{\HerwigPP}{\textsf{Herwig++}}
\newcommand{\Herwig}{\textsf{Herwig}}
\newcommand{\Matchbox}{\textsf{Matchbox}}

\newcommand{\powheg}{\textsc{Powheg}}
\newcommand{\powhegbox}{\textsc{PowhegBox}}

\begin{document}
\begin{titlepage}
\pubblock

\vfill
\Title{Top Quark Production and Decay in \Herwig~7.1}
\vfill
\Author{
  Johannes Bellm$^{1}$,
  Kyle Cormier$^{2}$,
  Stefan Gieseke$^{3}$,
  Simon Pl\"atzer$^{4}$,
  Christian Reuschle$^{5}$,
  Peter Richardson$^{6,7}$,
  Stephen Webster\support$\,^{7}$
}
\Address{\institute}

\vfill
\begin{Abstract}
  A summary of recent developments in the simulation of top
  quark production and decay in the \Herwig\ Monte Carlo event generator.  
\end{Abstract}
\vfill
\begin{Presented}
$10^{th}$ International Workshop on Top Quark Physics\\
Braga, Portugal, September 17--22, 2017
\end{Presented}
\vfill
\end{titlepage}
\def\thefootnote{\fnsymbol{footnote}}
\setcounter{footnote}{0}

\section{Introduction}

We review recent developments in the simulation of top quarks
in \Herwig~7.0~\cite{Bellm:2015jjp,Bahr:2008pv} and \Herwig~7.1~\cite{Bellm:2017bvx}.
We give an outline of relevant developments in the angular-ordered and
dipole showers in \Herwig, work on the choice of the shower-starting scale
for next-to-leading order (NLO) matched $pp\to t\bar{t}$ events and a new
multi-jet merging algorithm that has recently been implemented in \Herwig.

\section{Angular-Ordered Shower Developments}

In the angular-ordered shower~\cite{Gieseke:2003rz} each outgoing parton from
the hard process,
referred to as a shower progenitor, is selected and showered separately.
First the values of the splitting variables are determined for each splitting
in the shower from each progenitor. Using these values the 
kinematics of the partons in each splitting are reconstructed, starting 
from the final splitting in the shower from each progenitor
and working towards the hard process.
Through this process the partons in the hard process
gain an unphysical off-shell mass and we must perform a reshuffling of the
momenta of the particles in the event to restore energy-momentum conservation.

The default method for this reshuffling has changed between \HerwigPP, \Herwig~7.0 and
\Herwig~7.1 such that we now make more use of the colour information from the 
hard process in our treatment of recoils in the procedure.
The reader should refer to~\cite{herwig7manual} for a detailed description
of the changes. These developments in the reshuffling procedure
have been driven by effects seen in distributions of top quark observables, 
in particular the invariant mass of the $t\bar{t}$-pair, obtained for
\powhegbox~\cite{Nason:2004rx,Frixione:2007vw,Alioli:2010xd}
$pp\to t\bar{t}$ events showered with \HerwigPP.
Improvements in these results between \HerwigPP\ and \Herwig~7.1
are primarily due to the changes in the shower reconstruction.

\section{Dipole Shower Developments}

The dipole shower~\cite{Platzer:2009jq,Platzer:2011bc} has
undergone significant developments between \Herwig~7.0 and \Herwig~7.1.
The kinematics used to describe splittings off dipoles
consisting of an initial-state emitter and a massive final-state spectator
(massive IF dipoles) and a final-state emitter and a
final-state spectator, including a massive parton before or after the splitting,
(massive FF dipoles) have been completely reformulated. We have also revised the Jacobians
required for the evaluation of the shower kernels and Sudakov form factors for
dipole splittings involving massive partons. These changes were implemented in
\Herwig~7.1 and the reader should refer to~\cite{herwig7manual} for full details.

\begin{figure}[htb]
\centering
\includegraphics[width=0.45\textwidth]{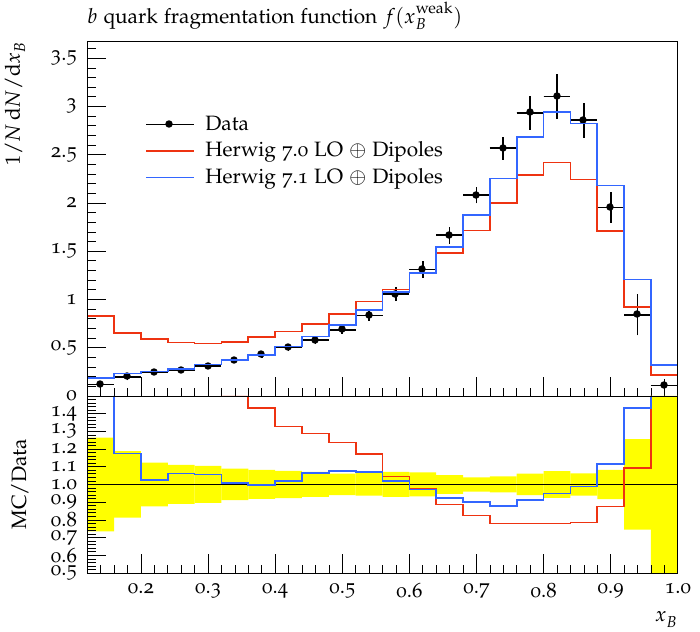}
\caption{The B-fragmentation function as measured by SLD~\cite{Abe:2002iq}.
  Predictions using the dipole shower in \Herwig~7.0 and the improved treatment in
  \Herwig~7.1 are shown.}
\label{plot:DipoleShower-BFrag}
\end{figure}

While the changes to the massive IF dipole are particularly important in
$t\bar{t}$-production, the improvements are most clearly seen in the prediction of
the B-fragmentation distribution at LEP, shown in Figure~\ref{plot:DipoleShower-BFrag},
which is highly dependent upon the massive FF dipole.
While this distribution was poorly described by the dipole shower in \Herwig~7.0,
the description is clearly improved in \Herwig~7.1.

In \Herwig~7.1 the dipole shower has been extended to include the showering of
top quarks in their decay.
We treat top quark decays in the dipole shower in the 
narrow-width approximation and we require that the momentum of each
top quark, set by the hard process and its showering, is conserved in its decay
and subsequent showering.
In addition, the NLO correction to the first emission off the top quark decay is
available using the built-in \powheg\ decay correction in
\Herwig~\cite{Richardson:2013nfo}. 

With these developments both parton showers in \Herwig\ can shower top quarks in
their production and decay at NLO in QCD. The production process can be
matched at NLO using either the subtractive-type or multiplicative-type
matching schemes available through the \Matchbox\ module~\cite{Platzer:2011bc}
in \Herwig.

\section{Shower Scale for NLO Matching in $pp\to t\bar{t}$}

In MC@NLO-type~\cite{Frixione:2002ik} events, referred to as subtractive-type matching,
$\oplus$, in our {\Matchbox}-specific terminology, we must choose the scale, $Q_\mathrm{shower}$,
from which we begin showering the hard process. In general the effects of the choice of
$Q_\mathrm{shower}$ are of a higher-order than the formal accuracy of the calculation,
therefore it should be considered as a shower uncertainty. The default setting in
\Herwig\ is $Q_\mathrm{shower} = \mu_\mathrm{F}$, where $\mu_\mathrm{F}$
is the factorisation scale. In \Herwig~7.1 we have introduced a new optional choice for
the shower-starting scale for use in $pp\to t\bar{t}$. We have introduced this scale
to allow the developers and users of \Herwig\ to investigate 
the effects of using an alternative functional form for this scale.
The new scale option, $\mu_\mathrm{opt}$, is,
\begin{equation}
  \mu_\mathrm{opt}^2 = \frac{1}{n_\mathrm{out}}
  \sum_{i=1}^{n_\mathrm{out}} m_{\mathrm{T},i}^2 \ ,
\end{equation}
where $n_\mathrm{out}$ is the number of particles outgoing from the hard process
and the transverse mass, $m_{\mathrm{T},i}$, of the $i$th outgoing particle
is given in terms of the mass, $m_i$, and transverse momentum,
$p_{\mathrm{T},i}$, of the particle by $m_{\mathrm{T},i} = \sqrt{m_i^2 + p_{\mathrm{T},i}^2}$\ .

\begin{figure}[htb]
  \centering
  \subfloat[] {
    \includegraphics[width=0.45\textwidth]{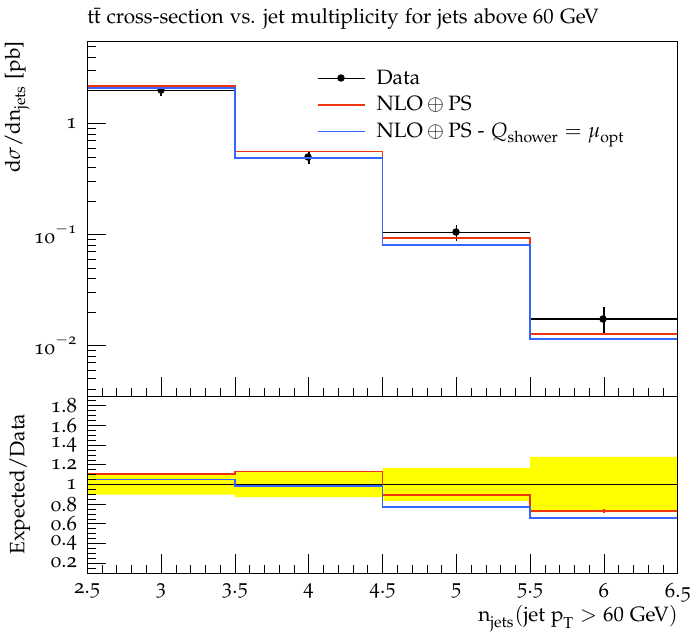}
  }
  \hfill
  \centering
  \subfloat[] {
    \includegraphics[width=0.45\textwidth]{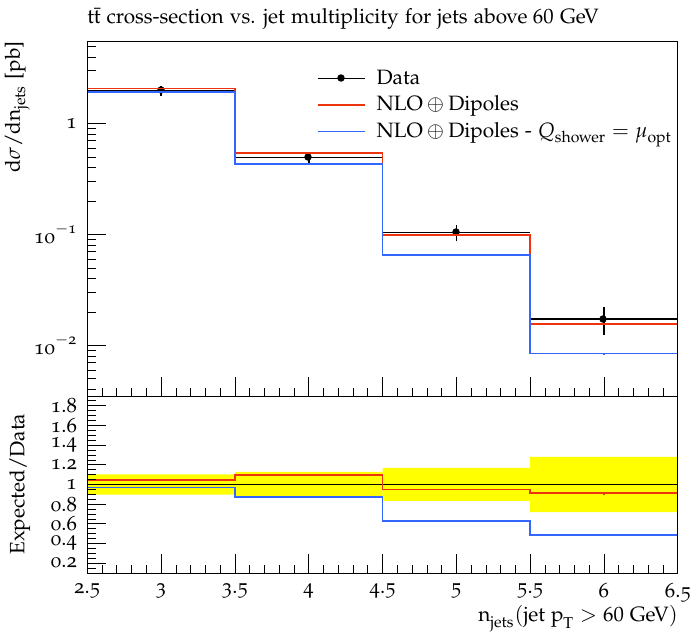}
  }
  \caption{The jet-multiplicity in $pp\to t\bar{t}$ at 7\,TeV as measured by
    ATLAS~\cite{Aad:2014iaa}
    and predictions using the angular-ordered (PS) and dipole showers, with and without
    the new optional shower-starting scale choice.}
  \label{plot:ShowerScale-Both}
\end{figure}

In Figure~\ref{plot:ShowerScale-Both} we show the jet multiplicity, $n_\mathrm{jets}$,
distribution in 7\,TeV $pp\to t\bar{t}$ events for jets with transverse momentum
greater than 60 GeV. The predictions have been produced with the factorisation scale
chosen to be the invariant mass of the $t\bar{t}$-pair.
Results from both showers, with and without the alternative
shower-starting scale, are presented. We see that the new scale choice produces
a decrease in all multiplicity bins, however the effects are more pronounced in the
dipole shower prediction. The reader should refer to~\cite{HerwigTopPaper}
for a more detailed discussion of this topic.

\section{NLO Multi-jet Merging}

A new NLO multi-jet merging algorithm based on the unitarised merging paradigm
has been introduced in \Herwig~7.1~\cite{Platzer:2012bs,Bellm:2017ktr}. 
This new implementation builds upon the existing \Matchbox\ infrastructure
in \Herwig\ and is currently available for merging with the dipole shower. 
For a given process one can merge the leading order matrix elements (MEs)
for additional jet multiplicities and apply the NLO corrections to these MEs
as required. In principle it is possible to merge an
abitrary number of jets, however in practice the multiplicity is limited
by the availability of the required MEs from external libraries and by the
computation time required to merge large numbers of additional jets.

\begin{figure}[htb]
\centering
\includegraphics[height=0.45\textwidth]{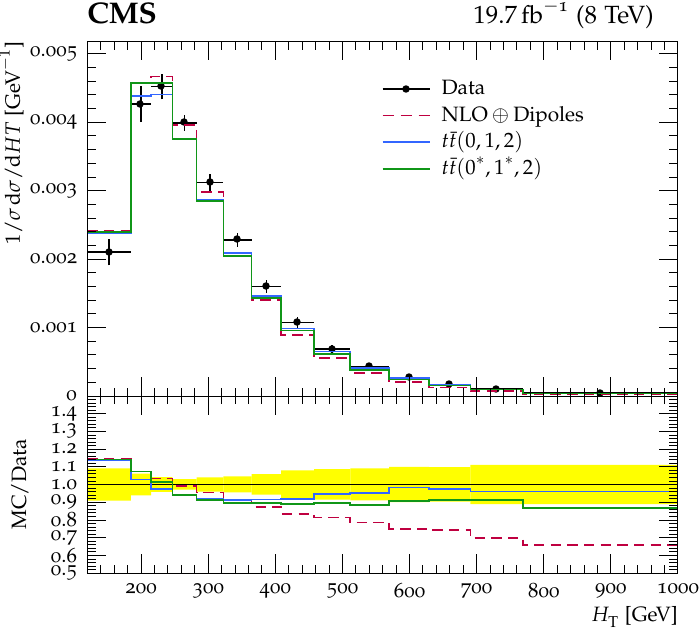}
\caption{The distribution of $H_\mathrm{T}$ in $pp\to t\bar{t}$ at 8\,TeV as measured by
  CMS~\cite{Khachatryan:2016oou} and predictions of the dipole shower in a NLO-matched
  sample and two multi-jet merged samples.}
\label{plot:Merging-HT}
\end{figure}

Several observables of interest in $pp\to t\bar{t}$ are not expected to be well-described
by NLO-matched samples. An example, shown in Figure~\ref{plot:Merging-HT},
is the $H_\mathrm{T}$ distribution, where $H_\mathrm{T}$ is the scalar sum of
the transverse momentum of all outgoing jets from each event. It is evident that
the NLO-matched result does a poor job of describing the data across much of the
distribution.
The results from two merged samples are also given. The $t\bar{t}(0,1,2)$ sample
merges tree-level MEs for $t\bar{t}$-production with 0-, 1- and 2-additional parton emissions,
while the $t\bar{t}(0^\ast,1^\ast,2)$ sample also includes the one-loop MEs for 
0- and 1-additional parton emissions. The results from the merged 
samples display an evident improvement over the NLO-matched result.

\Acknowledgements
SW is grateful to the organizers of the Top2017 conference for the opportunity
to present these new developments in \Herwig. 
SW acknowledges support from a STFC studentship.
KC is grateful to The University of Manchester for hosting him during part of this work.
This work was supported in part by the European Union as part of the FP7
Marie Curie Initial Training Network MCnetITN (PITN-GA-2012-315877).
In addition this work has received funding from the European Union's Horizon
2020 research and innovation programme as part of the Marie Skłodowska-Curie
Innovative Training Network MCnetITN3 (grant agreement no. 722104).

All contributors, hosting institutions and funding agencies who gave invaluable input,
feedback and support towards the developments described in this note, as part of the
work towards the release of \Herwig~7 and \Herwig~7.1, are acknowledged in
\cite{Bellm:2015jjp,Bellm:2017bvx}.

\end{document}